# Entanglement and final state of a black hole under general unitary transformation

Doyeol Ahn*

*Center for Quantum Information Processing*
*Department of Electrical and Computer Engineering, University of Seoul,*
*Jeonnong 90, Tongdaimun-ku, Seoul 130-743, Republic of Korea*

**Abstract**

Final state of a black hole is studied by taking into account a general unitary evolution of a black hole matter state. It is found that the mixedness decreases under general unitary transformation when the initial matter state was in mixed state and the mean fidelity at the evaporation is smaller than the fidelity of the quantum teleportation by the factor of inverse square of the number of states of a black hole. The change of entanglement Alice-Bob pair at evaporation is also studied by calculating the entanglement fidelity and eigenvalues of the partial transposed block density matrix. We assumed that Alice fell onto the black hole when the hole was formed while Bob is maintaining the safe distance from the black hole. It is found that the entanglement fidelity is inversely proportional to the square of the Hilbert space dimension N and the entanglement can survive the evaporation process.

*To whom correspondence should be addressed. E-mail: *dahn@uos.ac.kr*,
davidahn@hitel.net

## I. Introduction

Black hole information paradox has been a serious challenge to fundamental physics for over three decades [1-11]. Hawking [3] predicted that a process of black hole formation and evaporation is not unitary and the information consumed by a black hole is lost because the final state is represented by a density matrix rather than a pure quantum state in his model. On the other hand, string theory suggests that the formation and evaporation of black hole could be a unitary process [4]. Especially, Horowitz and Maldacena (HM) [4] proposed a model of black hole evaporation with a final-state boundary condition for the final state of an evaporating black hole at its singularity. The resulting projection onto a final state yields a pure quantum state outside of the black hole by a process resembling the quantum teleportation. The essential conjecture of HM proposal is to impose a unique final boundary condition at the black hole singularity such that no information is absorbed by the singularity. Recently, the author proved [8,10] this conjecture for the case of collapsing gravitational shell of the Schwarzschild black hole for both bosons [8] and fermions [10]. In previous work [8, 10, 11], our analysis is based on the assumption that the evolution of the black hole internal state is determined by the unitary transformation in the product form given by product form $U = S_1 \otimes S_2$ where $S_1 : H_M \to H_M$ and $S_2 : H_{in} \to H_{in}$ where $H_M$ is the Hilbert space of the initial matter state and $H_{in}$ is the Hilbert space which contains quantum states localized inside of the event horizon. Here, we generalize the previous approach by employing the general unitary transformation $U : H_M \otimes H_{in} \to H_M \otimes H_{in}$ with the normalization condition $\frac{1}{N}\sum_{j,k=0}^{N-1}|U_{jk}|^2 = 1$ to model the internal black hole state evolution and study the mixedness, mean fidelity of final state and the entanglement between Alice-Bob pair with Alice fallen onto the event horizon. Originally, Alice was entangled with an external observer Bob at the formation of a black hole. Bob is assumed to remain in the safe distance from the black hole through the process.

## II. Black hole final state, mean fidelity and entanglement

We assume that a quantum state of the collapsing matter belongs to a Hilbert space $H_M$ with dimension $N$ and $|\phi\rangle_M$ be the initial quantum state of a collapsing matter. The Hilbert space of fluctuations on the background spacetime for black hole formation and evaporation is separated into $H_{in}$ and $H_{out}$ which contain quantum states localized inside and outside the event horizon, respectively. The Unruh vacuum state $|\Phi_0\rangle_{in \otimes out}$ belonging to $H_{in} \otimes H_{out}$ in micro-canonical form is given by[4-6]:

$$|\Phi_0\rangle_{in \otimes out} = \frac{1}{\sqrt{N}} \sum_{l=0}^{N-1} |l\rangle_{in} \otimes |l\rangle_{out}, \tag{1}$$

where $\{|l\rangle_{in}\}$ and $\{|l\rangle_{out}\}$ are orthonormal bases for $H_{in}$ and $H_{out}$, respectively.

The final state is related to time "symmetric" formulation of quantum mechanics [11,12]. In this formulation, the conditional probability of the system to experience a chain of projection (or measurements) denoted by the operator $C(\alpha)$ when both initial and final states are fixed, is given by

$$P(\tilde{c}(\alpha) | \tilde{a}, \tilde{b}) = \frac{Tr(\rho_b C(\alpha) \rho_a C^\dagger(\alpha))}{\sum_\alpha Tr(\rho_b C(\alpha) \rho_a C^\dagger(\alpha))}, \tag{2}$$

where $\rho_a = |a\rangle\langle a|$ is the initial state at the past null infinity $J^-$, $\rho_b = |b\rangle\langle b|$ is the state at the future null infinity $J^+$ and $\alpha = (\alpha_1, \cdots, \alpha_n)$ at $(t_1, \cdots, t_n)$ is a sequence of alternatives. Here, $\tilde{a}, \tilde{b}, \tilde{c}(\alpha)$ are eigenvalues or sequence of eigenvalues. Let $\{|n\rangle\}$ be the basis of infinite dimensional Hilbert space $H$. We assume that the initial matter state which collapses to a black hole belongs to $H_M$. The initial state at the past null infinity $J^-$ (Fig. 1) is assumed to be the tensor product of the initial matter state in $H_M$ and the Unruh vacuum state [13-15] belonging to $H_{in} \otimes H_{out}$ where $H_{in}$ and $H_{out}$ denote Hilbert spaces which contain quantum states localized inside and outside the event horizon, respectively [8]. On the other hand, the final state at the future null infinity $J^+$ is assumed to be the tensor product state of an outgoing wave in $H_{out}$ and a Unruh vacuum state $|\Psi\rangle_{M \otimes in}$ belonging to $H_M \otimes H_{in}$. The author has shown that the ground state of a collapsing shell inside the Schwarzschild black hole can be obtained by the Bogoliubov transformation of the pair of creation and annihilation operators

belonging to $H_M$ and $H_{in}$ [8]. We consider only the chain of projections $C_{out}(\alpha)$ associated with an outside observer and assume that $\rho_a = \rho_M \otimes \rho_{in \otimes out}$ and $\rho_b = \rho_{M \otimes in} \otimes \frac{I_{out}}{\tilde{N}}$, where $\tilde{N} = \sum_\alpha Tr(\rho_b C(\alpha) \rho_a C^\dagger(\alpha))$. Here, $\rho_M = |\psi\rangle_M \langle\psi|$, $\rho_{in \otimes out} = |\Phi_0\rangle_{in \otimes out} \langle\Phi_0|$, and $\rho_{M \otimes in} = |\Psi\rangle_{M \otimes in} \langle\Psi|$. Then, the conditional probability becomes [11]

$$\begin{aligned} P(\tilde{c}(\alpha) | \tilde{a}, \tilde{b}) &= \frac{Tr(\rho_b C(\alpha) \rho_a C^\dagger(\alpha))}{\sum_\alpha Tr(\rho_b C(\alpha) \rho_a C^\dagger(\alpha))} \\ &= \frac{Tr(\rho_{M \otimes in} \otimes I_{out} C_{out}(\alpha) \rho_{in \otimes out} C_{out}^\dagger(\alpha))}{\tilde{N}^2} \\ &= \frac{Tr(\rho_{M \otimes in} \otimes I_{out} C_{out}(\alpha) \rho_M \otimes \rho_{in \otimes out} C_{out}^\dagger(\alpha))}{\tilde{N}^2} \\ &= \frac{Tr(C_{out}(\alpha) \rho_{M \otimes in} (\rho_M \otimes \rho_{in \otimes out}) C_{out}^\dagger(\alpha))}{\tilde{N}^2} \\ &= Tr(C_{out} |\tilde{\phi}\rangle\langle\tilde{\phi}| C_{out}^\dagger(\alpha)), \end{aligned} \qquad (2)$$

where

$$|\tilde{\phi}\rangle = \frac{1}{\tilde{N}} {}_{M \otimes in}\langle\Psi|\left(|\psi\rangle_M \otimes |\Phi_0\rangle_{in \otimes out}\right). \qquad (3)$$

In Eq. (3), the dual state ${}_{M \otimes in}\langle\Psi|$ can be regarded as a final-state boundary condition (FBC) at the future null infinity. The final-state boundary condition imposed at the singularity requires a maximally entangled quantum state in $H_M \otimes H_{in}$ which is called final boundary state and is given by [11]

$$\begin{aligned} {}_{M \otimes in}\langle\Psi| &= \frac{1}{\sqrt{N}} \sum_{l=0}^{N-1} {}_M\langle l| \otimes {}_{in}\langle l| U^\dagger = \frac{1}{\sqrt{N}} \sum_{j,k=0}^{N-1} {}_M\langle j| \otimes {}_{in}\langle k| U^*_{jk}, \\ U^*_{jk} &= \sqrt{N} \left({}_M\langle j| \otimes {}_{in}\langle k|\right) |\Psi\rangle_{M \otimes in} \end{aligned} \qquad (4)$$

where $U$ is a general unitary map $U : H_M \otimes H_{in} \to H_M \otimes H_{in}$ with the normalization condition $\frac{1}{N} \sum_{j,k=0}^{N-1} |U_{jk}|^2 = 1$.

The Horowitz-Maldacena mechanism of black hole evaporation under general unitary evolution is described by the following process:

$$_{M \otimes in}\langle \Psi |[|\phi\rangle_M \otimes |\Phi_0\rangle_{in \otimes out}] = P_{out,M}|\phi\rangle_M = |\tilde{\phi}\rangle_{out},$$

$$|\tilde{\phi}\rangle_{out} = \frac{1}{N} \sum_{l,m=0}^{N-1} |m\rangle_{out} U^*_{lm\,M}\langle l|\phi\rangle_M, \qquad (5)$$

$$P_{out,M} = \frac{1}{N} \sum_{l,m=0}^{N-1} |m\rangle_{out} U^*_{lm\,M}\langle l|,$$

where $P_{out,M}$ is a transfer operator which projects the state vector belonging to $H_M$ onto $H_{out}$. This indicates that the pure state evolves into a pure state and the expansion coefficient $U^*_{lm}$ dictates how the information about the collapsing matter is mixed together with the information in the in-falling Hawking radiation.

We first consider how the mixed state evolves and the mixedness, the measure of information content, changes under the general transformation. The mixed state is represented by the density operator of the form

$$\rho_M = \sum_n C_n |n\rangle\langle n|, \qquad (6)$$

with $\quad \sum_n C_n = 1. \qquad (7)$

A mixed state arises when the initial matter state was entangled with other quantum systems when the black hole was about to be formed and was taken as a partial trace over these other systems. There are several ways to quantify the amount of quantum information content. In this work, we employ the mixedness suggested by Brukner and Zeilinger [16] as an experimentally relevant measure of information. The mixedness is defined as

$$M_x = Tr(\rho_M^2) \qquad (8)$$

for a qubit system.

From the transfer theorem [11], the final density operator at evaporation is obtain by

$$\rho_{out} = {}_{M \otimes in}\langle \Psi |(\rho_M \otimes |\Phi_0\rangle_{in \otimes out}\langle \Phi_0|)|\Psi\rangle_{M \otimes in}$$
$$= P_{out,M} \rho_M P^\dagger_{M,out}. \qquad (9)$$

The measure of mixedness is then given by

$$Tr(\rho_{out}^2) = Tr(W \rho_M W \rho_M), \qquad (10)$$

where

$$W = P_{out,M}^{\dagger} P_{out,M} . \tag{11}$$

For all bounded linear operators $X$ and density operator $T$, we have the following inequality [17]

$$|Tr(XT)| \leq \|X\| \|T\|_1 , \tag{12}$$

where $\|\cdot\|$ is an operator norm and $\|\cdot\|_1$ is a trace norm. Then from equations (11) and (12), we obtain

$$\begin{aligned} Tr(\rho_{out}^2) &= Tr(W\rho_M W\rho_M) \\ &\leq \|W\| Tr(\rho_M W\rho_M) \\ &= \|W\| Tr(W\rho_M^2) \\ &\leq \|W\|^2 Tr(\rho_M^2) \\ &\leq Tr(\rho_M^2) \end{aligned} \tag{13}$$

since $\sup W = 1$. This suggests that the measure of mixedness is expected to decrease or state become more mixed under black hole evaporation.

The robustness of the retrieval or escape of quantum information is related to the fidelity at the evaporation [7]. This fidelity is the probability that the final quantum state would pass a test checking whether it agreed with the initial quantum state. The probability that the observer outside of the black hole detects outgoing Hawking radiation at evaporation $|\tilde{\phi}\rangle_{out}$ for given arbitrary quantum matter state $|\phi\rangle_M$ is $_{out}\langle\tilde{\phi}|\tilde{\phi}\rangle_{out}$. The overlap of the normalized final state with the initial matter state is given by $|\langle\tilde{\phi}|\phi\rangle|^2 / \langle\tilde{\phi}|\tilde{\phi}\rangle$. The integration of this expression over all possible input states $|\phi\rangle_M$ with the weights $_{out}\langle\tilde{\phi}|\tilde{\phi}\rangle_{out}$ yields the expression for the mean fidelity of evaporation

$$\begin{aligned} \overline{F}_{EV} &= \int d\phi \left| \langle\tilde{\phi}|\phi\rangle \right|^2 \\ &= \frac{1}{N^2} \sum_{j,k,m,l} U_{jk}^* U_{lm} \langle k|\hat{M}_{lm}|l\rangle \end{aligned} \tag{14}$$

where the operators $\hat{M}_{lm}$ are given by the following integrals over the space of pure state $|\phi\rangle$ [18]

$$\hat{M}_{lm} = \int d\phi \langle m|\phi\rangle\langle\phi|l\rangle|\phi\rangle\langle\phi|. \tag{15}$$

Here $\hat{I}$ denotes an identity operator and the integral $\int d\phi$ over the space of pure state is performed using the canonical measure invariant with respect to unitary transformations of the states $|\phi\rangle$ [18,19]. In eqs. (14) and (15), we assume the arbitrary pure state in the N-dimensional Hilbert space given by

$$|\phi\rangle = \begin{pmatrix} e^{i\xi}\cos\theta \\ \sin\theta\cos\varphi \\ z_2 \sin\theta\cos\varphi \\ \vdots \\ z_{N-1}\sin\theta\cos\varphi \end{pmatrix}, \tag{16}$$

where $\xi, \theta$ and $\varphi$ are integral variables and $z_2, z_3, \cdots, z_{N-1}$ are parameters such that

$$|z_2|^2 + \cdots + |z_{N-1}|^2 = 1. \tag{17}$$

The invariant volume element is given by

$$d\phi = \frac{(N-1)!}{\pi^{N-1}}(\sin\theta)^{2N-2}(\sin\varphi)^{2N-5}d(\sin\theta)d(\sin\varphi)d\xi dS_{2N-5} \tag{18}$$

where $dS_{2N-5}$ is the volume element of the unit sphere $S_{2N-5}$. The volume of the unit sphere $S_d$ is given by $S_d = \dfrac{2\pi^{(d+1)/2}}{\Gamma(\frac{d+1}{2})}$. The invariant volume element $d\phi$ is the differential volume in which two vectors selected at random are separated by the Hilbert space angle between $(\xi, \theta, \varphi)$ and $(\xi+d\xi, \theta+d\theta, \varphi+d\varphi)$ occupy. As an example,

$$\langle 0|M_{00}|0\rangle = \frac{(N-1)!}{\pi^{N-1}}\int \cos^4\theta(\sin\theta)^{2N-3}(\sin\varphi)^{2N-5}d(\sin\theta)d(\sin\varphi)d\xi dS_{2N-5}$$
$$= \frac{2}{N(N+1)}, \tag{19}$$

and

$$\langle 1|M_{00}|1\rangle$$
$$=\frac{(N-1)!}{\pi^{N-1}}\int\cos^2\theta\left(1-\sin^2\theta\right)\left(1-\sin^2\phi\right)(\sin\theta)^{2N-3}(\sin\varphi)^{2N-5}d(\sin\theta)d(\sin\varphi)d\xi dS_{2N-5}$$
$$=\frac{1}{N(N+1)}.$$

(20)

Equations (19) and (20) can be generalized to yield,

$$\hat{M}_{lm}=\int d\phi\langle m|\phi\rangle\langle\phi|l\rangle|\phi\rangle\langle\phi|=\frac{1}{N(N+1)}\left(\delta_{lm}\hat{I}+|l\rangle\langle m|\right). \tag{21}$$

From equations (14) and (21), we obtain

$$\begin{aligned}\overline{F}_{EV}&=\frac{1}{N^2}\sum_{j,k,l,m}U_{jk}^*U_{lm}\langle k|\frac{\delta_{lm}\hat{I}+(|l\rangle\langle m|)}{N(N+1)}|j\rangle\\ &=\frac{1}{N^3(N+1)}\left(\sum_{j,k}|U_{jk}|^2+\left|\sum_{l}U_{ll}\right|^2\right)\\ &=\frac{1}{N^2}\left(\frac{1}{N+1}+\frac{1}{N(N+1)}\left|\sum_{l}U_{ll}\right|^2\right)\\ &\leq\frac{1}{N^2}\left(\frac{1+\left|\sum_{k}\lambda_k\right|^2}{N+1}\right)=\frac{\overline{F}_{QT}}{N^2}.\end{aligned} \tag{22}$$

Here $\overline{F}_{QT}=\frac{1}{N+1}\left[1+\left(\sum_{k}\lambda_k\right)^2\right]$ denotes the maximum average fidelity for the corresponding case of quantum teleportation of arbitrary pure state with $N$ levels and $\lambda_k$ is the Schmidt coefficient obtained from the singular value decomposition. In equation (19), we have used the singular value decomposition [20] $U=VDW$ where $D=\left(\lambda_i\delta_{ij}\right)$ is a diagonal matrix with real, non-negative entries, and $V$, $W$ are unitary matrices. Then, we obtain

$$\begin{aligned}\sum_{l}U_{ll}&=\sum_{k,l}\lambda_k V_{lk}W_{kl}\\ &\leq\sum_{k}\lambda_k\sum_{l,m}V_{lm}W_{lm}^*\leq\sum_{k}\lambda_k\sqrt{\sum_{l,m}|V_{lm}|^2}\sqrt{\sum_{l,m}|W_{lm}|^2}=N\sum_{k}\lambda_k.\end{aligned} \tag{23}$$

This result suggests that the mean fidelity at the evaporation is smaller than the fidelity of the quantum teleportation by the factor of inverse square of the number of states of a black hole, $N$. The final state condition at the evaporation resembles the generalized measurement in quantum teleportation to some extent. However, the resemblance ends here because during the black hole evaporation there is no way to communicate the information of $U_{jk}$ through the event horizon to an outside observer of a black hole. As a result, the decoding unitary process cannot be included in the black hole evaporation in the final state formalism, unlike the case of a conventional quantum teleportation. Equations (14) and (22) indicate that the final state projection selects one outcome (or operation) out of possible $N^2$ outcomes. If the receiver has the decoding unitary, in this case he does not, one would be able to retrieve the original quantum state. This fact could possibly explain the reduction of the fidelity in the case of the evaporation given in eq. (22).

We now consider the entanglement between Alice who fell onto the event horizon Bob who is maintaining the safe distance from the black hole. Alice was entangled with an external observer Bob at the formation of a black hole in the past null infinity $J^-$. The Unruh vacuum state is evolved unitarily from the vacuum state $|0_-\rangle$ defined in the far past [21,22]. If we define $a$ and $V$ as the annihilation operator of a particle in the far past and the unitary operator for the evolution of the vacuum state $|0_-\rangle$ due to the black hole formation, respectively, then we have $|\Phi_0\rangle_{in \otimes out} = V|0_-\rangle$ with $a|0_-\rangle = 0$. We also consider the excited state $|1_-\rangle$ with a finite energy at the past null infinity which is defined by $|1_-\rangle = a^\dagger |0_-\rangle$ [21]. Then the unitary evolution of this one-particle state at the future null infinity is given by [21]

$$V|1_-\rangle = Va^\dagger |0_-\rangle = Va^\dagger V^{-1}|\Phi_0\rangle_{in \otimes out} = \sqrt{\frac{2}{N(N+1)}} \sum_{l=0}^{N-1} \sqrt{l+1} |l\rangle_{in} \otimes |l+1\rangle_{out}. \quad (24)$$

The Unruh excited state at the future null infinity is then defined by $|\Phi_1\rangle_{in \otimes out} = V|1_-\rangle$ [21,22]. The Horowitz-Maldacena mechanism of black hole evaporation for the Unruh excited state is $_{M \otimes in}\langle \Psi |[|\phi\rangle_M \otimes |\Phi_1\rangle_{in \otimes out}] = \sqrt{\frac{2}{N+1}} b_{out}^\dagger P_{out,M}|\phi\rangle_M$ where $b_{out}^\dagger$ is the particle creation operator acting on $H_{out}$.

Suppose that Alice, a physicist of a highly advanced civilization in the distant future, just arrived at the vicinity of the black hole formation and fell into a black hole. On the other hand, Bob who originally is entangled with Alice, is located far from the black hole formation site. The maximally entangled state shared by Alice and Bob in the far past is given by $\frac{1}{\sqrt{2}}(|0_A\rangle \otimes |1_A\rangle + |1_A\rangle \otimes |0_B\rangle)$ where the subscripts $A$ and $B$ denote Alice and Bob, respectively. When a black hole is formed, Alice's state is affected by the unitary evolution operator $V$ while Bob's state remains intact because curvature nearby Bob is not affected by the black hole formation. When Alice fell into a black hole, the entangled state shared by Alice and Bob becomes $\frac{1}{\sqrt{2}}(|\Phi_{A;0}\rangle_{in\otimes out} \otimes |1_B\rangle + |\Phi_{A;1}\rangle_{in\otimes out} \otimes |0_B\rangle)$ and the total initial state is given by

$$|X\rangle = \frac{1}{\sqrt{2}}(|\Phi_{A;0}\rangle_{in\otimes out} \otimes |1_B\rangle + |\Phi_{A;1}\rangle_{in\otimes out} \otimes |0_B\rangle) \otimes |\phi\rangle_M,$$
(25)

$$|\Phi_{A;0}\rangle_{in\otimes out} = \frac{1}{\sqrt{N}} \sum_{l=0}^{N-1} |l\rangle_{A;in} \otimes |l\rangle_{A;out}, \quad |\Phi_{A;1}\rangle = \sqrt{\frac{2}{N(N+1)}} \sum_{l=0}^{N-1} \sqrt{l+1} |l\rangle_{A;in} \otimes |l+1\rangle_{A;out}$$

where the subscript A denotes Alice.

The quantum state at the evaporation by the Horowitz-Maldacena mechanism is obtained by the partial inner product [11]

$$|\psi\rangle_{AB} = {}_{M\otimes in}\langle\Psi|X\rangle = \sum_{m=0}^{N-1} f_m \left[\frac{1}{\sqrt{N}} |m\rangle_{A;out} \otimes |1\rangle_B + \sqrt{\frac{m+1}{N+1}} |m+1\rangle_{A;out} \otimes |0\rangle_B\right],$$
(26)
$$f_m = \frac{1}{N} \sum_{l=0}^{N-1} U^*_{lm\ M}\langle l|\phi\rangle_M.$$

The fidelity of the evaporation process $F_e$ is given by

$$F_e = Tr[|\psi\rangle_{AB}\langle\psi|Tr_{M,A;in}(|X\rangle\langle X|)] = \frac{1}{N(N+1)^2} \sum_{m=0}^{N-1} |f_m|^2 \left\{(n+1)^2 + \frac{(n+1)}{N+1} + \frac{1}{4(N+1)^2}\right\}.$$
(27)

We call $F_e$ the entanglement fidelity which is a measure of the quantum information preserved through the evaporation process.

Since $|f_m|^2 = \frac{1}{N^M} \langle \phi |[_{in}\langle m|\Psi\rangle_{M\otimes in} \langle \Psi|m\rangle_{in}]|\phi\rangle_M \approx \frac{1}{N^2}$, the entanglement fidelity is approximated by $F_e \approx \frac{13}{12N^2}$.

We also consider the change of entanglement between Alice and Bob through the evaporation process. The density operator for the final state becomes

$$\rho_{AB} = |\psi\rangle_{AB}\langle\psi|$$
$$= \sum_{m=0}^{N-1} (\rho_{AB})_m$$
$$= \sum_{m=0}^{N-1} |f_m|^2 \left\{ \frac{1}{2} |m1\rangle\langle m1| + \sqrt{\frac{m+1}{2(N+1)}} \left( |(m+1)0\rangle\langle m1| + |m1\rangle\langle(m+1)0| \right) + \frac{m+1}{N+1} |(m+1)0\rangle\langle(m+1)0| \right\}.$$

(28)

The matrix representation of $\rho_{AB}$ in the $[|m0\rangle, |m1\rangle, |(m+1)0\rangle, |(m+1)1\rangle]$ block $\rho_m^{AB}$ is given by

$$\rho_m^{AB} = \begin{pmatrix} \frac{m|f_{m-1}|^2}{N+1} & 0 & 0 & 0 \\ 0 & \frac{|f_m|^2}{2} & \sqrt{\frac{m+1}{2(N+1)}} |f_m|^2 & 0 \\ 0 & \sqrt{\frac{m+1}{2(N+1)}} |f_m|^2 & \frac{(m+1)|f_m|^2}{N+1} & 0 \\ 0 & 0 & 0 & \frac{|f_{m+1}|^2}{2} \end{pmatrix}. \quad (29)$$

Then the eigenvalues of the partial transposed density matrix are obtained as

$$\lambda = \frac{|f_m|^2}{2}, \frac{(m+1)|f_m|^2}{N+1}, \frac{1}{2}\left( \frac{|f_{m+1}|^2}{2} + \frac{m|f_{m-1}|^2}{N+1} \pm \sqrt{\left( \frac{|f_{m+1}|^2}{2} - \frac{m|f_{m-1}|^2}{N+1} \right) + \frac{2(m+1)|f_m|^4}{(N+1)}} \right). \quad (30)$$

Above result shows that one eigenvalue is always negative indicating there is an entanglement between Alice and Bob [23], provided

$$|f_m|^2 \geq \sqrt{\frac{m}{m+1}} |f_{m-1} f_{m+1}|. \quad (31)$$

Equation (31) should be considered as the constraint on the general unitary transformation for the survival of an entanglement between Alice and Bob.

## III. Summary

The results of this work suggests if we model the black hole evaporation by the final-state projection, the mixedness decreases under general unitary transformation when the initial matter state was in mixed state and the mean fidelity at the evaporation is smaller than the fidelity of the quantum teleportation by the factor of inverse square of the number of states of a black hole. We also considered the entanglement between Alice who fell onto the black hole when the hole is formed and Bob who remains in the safe distance from the black hole. Entanglement fidelity at the evaporation is found to the proportional to the inverse square of the number of black hole states as in the case of the mean fidelity. However, we found that the entanglement could survive the evaporation process.

**Acknowledgments**

This work was supported by the University of Seoul through the 2008 University Research grant.

**Figure legends**

**Fig. 1.** The Kruskal exptention of the Schwarzschild spacetime [19, 20]. In region *I*, null asymptotes $H_+$ and $H_-$ act as futue and past event horizons, respectively. The boundary lines labelled $J^+$ and $J^-$ are fute and past null infinities, respectively, and $i^o$ is the spacelike infinity.

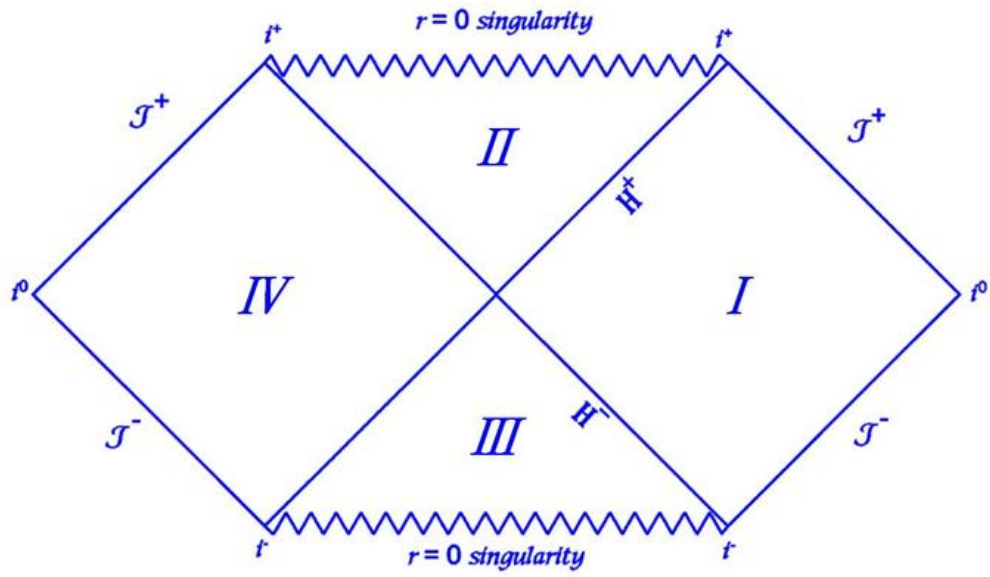

Fig. 1